# Automatic Detection of Search Tactic in Individual Information Seeking: A Hidden Markov Model Approach

**Shuguang Han**
University of Pittsburgh
shh69@pitt.edu

**Zhen Yue**
University of Pittsburgh
zhy18@pitt.edu

**Daqing He**
University of Pittsburgh
ah44@pitt.edu

## Abstract

Information seeking process is an important topic in information seeking behavior research. Both qualitative and empirical methods have been adopted in analyzing information seeking processes, with major focus on uncovering the latent search tactics behind user behaviors. Most of the existing works require defining search tactics in advance and coding data manually. Among the few works that can recognize search tactics automatically, they missed making sense of those tactics. In this paper, we proposed using an automatic technique, i.e. the Hidden Markov Model (HMM), to explicitly model the search tactics. HMM results show that the identified search tactics of individual information seeking behaviors are consistent with Marchionini's Information seeking process model. With the advantages of showing the connections between search tactics and search actions and the transitions among search tactics, we argue that HMM is a useful tool to investigate information seeking process, or at least it provides a feasible way to analyze large scale dataset.

## Background & Motivation

Information seeking process (ISP) is one of the major areas in information seeking behavior research. There are several well-known models for describing individual's information seeking process. Both Kuhlthau's [1] and Ellis'[2] models present holistic views of information seeking from the initiation stage to the ending stage, whereas Marchionini's model [3] consists of eight stages and it focuses on describing possible transitions among each of them. In addition to the above-mentioned macro-level investigations of search processes focusing on qualitative constructs of stages and context in ISP, there are also several empirical studies that look into micro-level search actions in ISP.

Some studies examined search processes through the actions in the user logs. For example, Holscher and Strube [4] compared action sequences between Internet experts and newbies. Chen and Cooper [5, 6] used both stochastic model and clustering techniques to examine search tactics in a Web-based library catalog. The problem of those approaches is that they missed explaining user intentions behind user actions. Xie and Joo [7] raised the importance of investigating transitions of search tactics as a means of examining search processes. They manually coded the transaction logs using a predefined scheme of search tactics. Then, a five-order Markov chain was adopted to find the common search tactics in user's behavior sequence.

Based on the literature review, we can see that investigation on information seeking process in individual user is an active research topic. Particularly, search tactics had been recognized as a mean of investigating search processes. However, most of previous researches either focus on a global qualitative analysis of search stages, or highly rely on manually coding of users logs. The manually coding of user actions is difficult to be expanded or used in a different or large-scale dataset. Although automatic

Acknowledgement: This work was partially supported by the National Science Foundation under Grant No. 0704628 and IIS-1052773.
Han, S., Yue, Z., & Daqing H. (2013). Automatic detection of search tactic in individual information seeking: A Hidden Markov Model approach. *iConference 2013 Proceedings* (pp. 712-716). doi:10.9776/13330

methods have been explored, they usually missed showing the connections between search tactics and search actions. The identified search tactics are simply the aggregation of certain user actions while the connection between user action and search tactic is missing. In this paper, by treating the sequence of user behaviors as a Markov chain, we modeled users' search tactics as hidden variables. Then, a HMM algorithm is used to identify those hidden search tactics. Each search tactic is then represented by a probability distribution over user actions, i.e. the probability of each search tactic generates each user action. Another outcome of HMM is the transition probabilities among the identified tactics, i.e. the probability of each search tactic transfers to another search tactic (including itself).

## Automatic Detection of Search Tactics

### Model Search Tactics

The HMM model for search tactic identification is depicted in Figure 1. Suppose we have a sequence of user actions from O1 to ON. We model search tactics as hidden states and assume each action is generated by one search tactic, from T1 to TN. Assume that we have R different types of user actions (from A1 to AR) and M different types of search tactic (from S1 to SM). Each search tactic Ti (1≤i≤N) is one of Si (1≤i≤M), and each user action Oi (1≤i≤N) is one of Ai (1≤i≤R). A HMM has several parameters: the number of user actions N, the number of user action types R, the number of search tactics M, the transition probabilities among search tactics, i.e. the probability between Si (1≤i≤M) and Sj (1≤j≤M) , and the emission probabilities from Si (1≤i≤M) to Ai (1≤i≤R). By only defining the M, the Baum-Welch algorithm [11] could be used for parameter estimation.

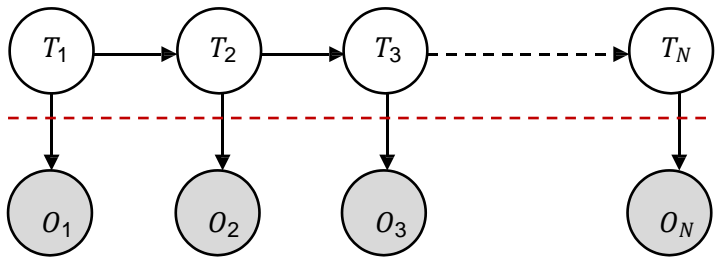

*Figure 1*: A Hidden Markov Model for Search Tactics

### Model Selection

It is still an open issue in determining the M in a HMM. A large M will increase the sequence likelihood because more parameters can describe data more precisely, but with a high risk of being over-fitting. On the other hand, a small M may be insufficient to describe the data. Choosing an optimal M is a model selection problem and Bayesian information criterion (BIC) [12] is usually adopted.

In order to use the BIC, we need to calculate the number of parameters NP. Since we have M search tactics and T actions, so there will be M*M transition probabilities and M*T emission probabilities. In HMM, we also need to define the prior probabilities of each search tactic, the number of which is M. Considering the constraint on the summation of probabilities equals to 1, the NP = M*(M-1)+M*(T-1)+(M-1). Suppose the sample size is S and the data likelihood is L. Then, the BIC is defined in Eq. (1). A large log(L) and less parameter NP are preferred. The smaller of BIC, the more preferred the model is.

$$BIC = -2 \times log(L) + log(S) \times NP \qquad \text{Eq. (1)}$$

## User Study and Dataset

In order to obtain the real data for users' information seeking behavior, we conducted a user study using a web search system that we built [8]. Seven students from the University of Pittsburgh who use computers on a daily basis were recruited to participate the study. Each participant was required to complete two search tasks with one on academic topic [9] and other one on leisure topic [10]. The goal of each task is to collect as much relevant information as possible.

Through analyzing the search logs obtained, we identified five typical types of search actions which are Query, View, Save, Workspace, Topic (See Table 1). The user study gives us a real dataset that consists of fourteen different behavior sequences (two tasks × seven participants).

Table 1: *User search actions*

| Actions | Descriptions |
|---|---|
| **Query (Q)** | A user issues a query or clicks a query from search history |
| **View (V)** | A user click on a result in the returned result list |
| **Save (S)** | A user saves a snippet or bookmarks a webpage |
| **Workspace (W)** | A user clicks, edits or comments on an item saved in the workspace |
| **Topic (T)** | A user clicks on the topic statement for view or leaves comments |

## Result and Discussion

### HMM Results

The BIC evaluation of our HMM model indicated that the optimal number of hidden states is 5 (See Figure 2), so our following analysis have all been based on M=5.

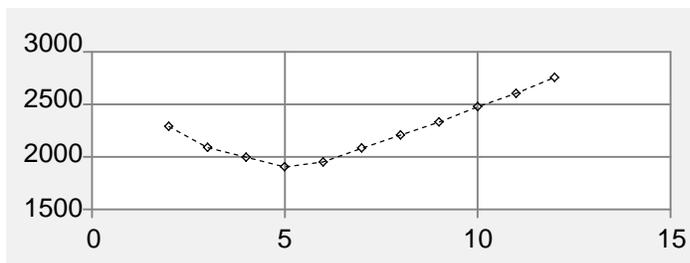

*Figure 2*: The BIC Evaluation

The emission probabilities are shown in Table 2. We remove the probabilities that are smaller than 0.05 for better visualization. Table 2 shows association between search tactics and user actions. For example, S1 refers to a search tactic that is mostly represented by Query, S3 refers to Save, S4 refers to Workspace, and S5 refers is represented by both Workspace and Topic. It is also interesting to observe that both S2 and S3 refer to View. To reveal the differences between S2 and S3, we looked at their transition probabilities.

Figure 3 visualizes the transition probabilities between all the search tactics. Each cell denotes a transition probability from the search tactic in the row to the search tactic in the column. The darker the cell is, the bigger the transition probability. The visualization is shown in Figure 3. Now it is clear that S2 tends to transmit to S3 while S3 tends to transmit to S4. It is probably that S2 is associated with quick scanning the results whereas S3 is more carefully checking results that lead to finding and saving relevant documents.

Table 2: *Search Tactics and Emission Probability*

|    | Q    | V    | S    | W    | T    |
|----|------|------|------|------|------|
| S1 | 0.92 |      |      |      | 0.06 |
| S2 |      | 0.97 |      |      |      |
| S3 |      | 0.98 |      |      |      |
| S4 |      |      | 0.97 |      |      |
| S5 |      |      |      | 0.67 | 0.32 |

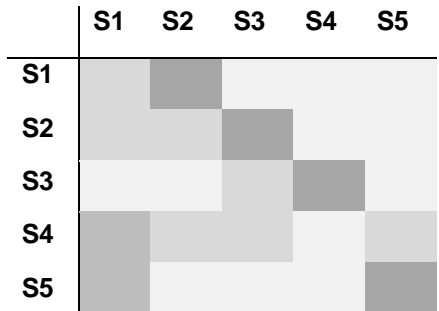

*Figure 3*: Transition of Search Tactics

**Comparison with Marchionini's ISP model**

Marchionini's ISP model is one of the well-known ISP models. It provides a clear description of user search behaviors in digital environment. His model divides user's information seeking process into eight sub-processes, and provides the transition possibilities among those sub-processes.

Because of the relative similarity between Machionini's model and our HMM model, we directly compare these two models. Some of Machionini's sub-processes are mapped jointly to one HMM search tactic. For example, the sub-process "*select source*" is predefined in our system because we use Google results by default. "*formulate query*", as its defined in the model is a cognitive work, which is difficult to be recorded. Therefore, we treat the "*select source*", "*formulate query*" and "*execute search*" as an integrative process called "*query*".

The comparison results are shown in Table 3. The search tactics obtained from HMM model can reasonably been mapped to sub-processes in Marchionini's ISP model. According to the ISP model, the main transition path is S5 → S1 → (S2 → S3) → S4, which almost perfectly matches to the darkest areas shown in Figure 3. Our HMM model mark the one "examining results" sub-process in the ISP model as two separate search tactics S2 and S3.

Table 3: *Mapping from sub-process to HMM patterns*

| Sub-processes      | Patterns |
|--------------------|----------|
| Define Problem     | S5       |
| Select Source      | S1       |
| Formulate Query    |          |
| Execute Query      |          |
| Examine Results    | S2, S3   |
| Extract Information| S4       |
| Reflect/Iterate/Stop | S5     |

## Conclusion

Search tactics have been recognized as a way of investigating information seeking processes. However, most of previous works are either based on predefined search tactics or simply aggregation of sequential user actions. In this paper, we adopted a Hidden Markov Model (HMM) approach to solve these two problems. By introducing the search tactics as hidden variables HMM can build connections between search tactics and search actions. Our result shows that HMM also produce a reasonable performance in our experiment because the result shows a considerable agreement with Marchionini's ISP model. HMM is a standard machine learning algorithm, which can easily be applied in a different, or a very large dataset.